# Gland Segmentation in Histopathology Images Using Deep Networks and Handcrafted Features

Safiyeh Rezaei, Ali Emami, Hamidreza Zarrabi, Shima Rafiei, Kayvan Najarian, Nader Karimi, Shadrokh Samavi, S.M.Reza Soroushmehr

*Abstract—* Histopathology images contain essential information for medical diagnosis and prognosis of cancerous disease. Segmentation of glands in histopathology images is a primary step for analysis and diagnosis of an unhealthy patient. Due to the widespread application and the great success of deep neural networks in intelligent medical diagnosis and histopathology, we propose a modified version of LinkNet for gland segmentation and recognition of malignant cases. We show that using specific handcrafted features such as invariant local binary pattern drastically improves the system performance. The experimental results demonstrate the competency of the proposed system against the state-of-the-art methods. We achieved the best results in testing on section B images of the Warwick-QU dataset and obtained comparable results on section A images.

## I. INTRODUCTION

Histopathological analysis is an essential process for cancer detection. Gland segmentation in histopathology images is a critical task in cancer diagnosis and prognosis. Every gland contains lumen, cytoplasm and epithelial layers. Traditionally, pathologists used to segment the glands manually, which was an inaccurate and time-consuming process, especially in large scales. Structure of glands is corrupted in malignant cases. While benign cases generally have circular structures, malignant glands demonstrate irregular shapes. Therefore, the automatic segmentation of malignant cases is challenging compared to benign structures. Furthermore, glands that stick together must be segmented and recognized as separate glands. To automate the segmentation of histopathology images with the problems mentioned above, machine vision researchers have proposed various algorithms for the automatic segmentation.

Some of the researches propose to use structural information of glands for segmentation [1], [2]. Gunduz-Demir *et al.* [1] decomposed image into a set of nucleus and lumen objects and used object-graph to estimate nucleus object edges. Then lumen objects were separated into two classes based on their textural properties: lumen objects inside a glandular region and outside of it. Lumen objects inside the glands were determined as initial gland seeds. Then gland regions were identified by growing the initial seed until they hit the graph edges. In the other work, Paul *et al.* [2] used informative morphological scale space to segment glands. They showed that the red channel of RGB has more information than other channels in histopathology images and designed a filter for the preservation of edges and segmentation of epithelial layers in the red channel.

Segmentation based on the structure of glands is faster compared to segmentation based on a neural network [2]. However, deep networks surpass the former systems in terms of accuracy. Since structures of glands are different in various types of diseases, hence in histopathology images, structural information is insufficient for segmentation [3]. Some recent works use neural networks and deep features alongside handcrafted features for segmentation of the glands [4], [5], [6]. Chen *et al.* [3] proposed a deep contour-aware network to segment contours and gland objects simultaneously. Thresholding the output probability maps attained final segmentation of glands. Li *et al.* [4] combined handcrafted features with fine-tuned deep features to train an SVM classifier. The classifier distinguishes gland and none gland windows. Then posterior probability at the center of each window is calculated and is thresholded to form segmentation results. Manivannan *et al.* [5], [6] extract raw-patches and features such as SIFT and multi-resolution Local Binary Patterns (LBP) from sliding windows of the image. Then, K-means is applied for clustering these windows to 30 groups based on the proposed features. Eventually, the clusters are classified by SVM. In the test phase, for each window, a weighted average of the patch probabilities provides the overall probability map and generates the final segmentation result. They improved their work [6] by adding deep features to their structure extracted from a pre-trained fully convolutional neural network. Xu *et al.* [7] use a convolutional neural network (CNN) and multichannel learning with the region and boundary cues to tackle the gland segmentation problem.

*Hematoxylin and Eosin* (*H&E*) is a staining technique in histology [12], which enhances the visibility of spatial structures and makes color information more distinguishable. For better use of color information, stain decomposition can be used in histopathologic images. When *H&E* stains are mixed up in an area, we get blurred color information. Hence, original

S. Rezaei, S. Rafiei, H.R. Zarrabi, N. Karimi are the Department of Electrical and Computer Engineering, Isfahan University of Technology, Isfahan 84156-83111, Iran.
K. Najarian and S.M.R. Soroushmehrs are with the Department of Computational Medicine and Bioinformatics and also the Michigan Center for Integrative Research in Critical Care, University of Michigan, Ann Arbor, 48109 U.S.A (e-mail {kayvan, ssoroush}@umich.edu).
S. Samavi is with the Department of Electrical and Computer Engineering, Isfahan University of Technology, Isfahan 84156-83111, Iran and also with the Department of Emergency Medicine, University of Michigan, Ann Arbor, 48109 U.S.A.
A. Emami is with the Department of Electrical and Computer Engineering, Isfahan University of Technology, Isfahan 84156-83111, Iran and also with the Department of Information Technology and Elect. Engineering, University of Queensland, QLD 4072, Australia.

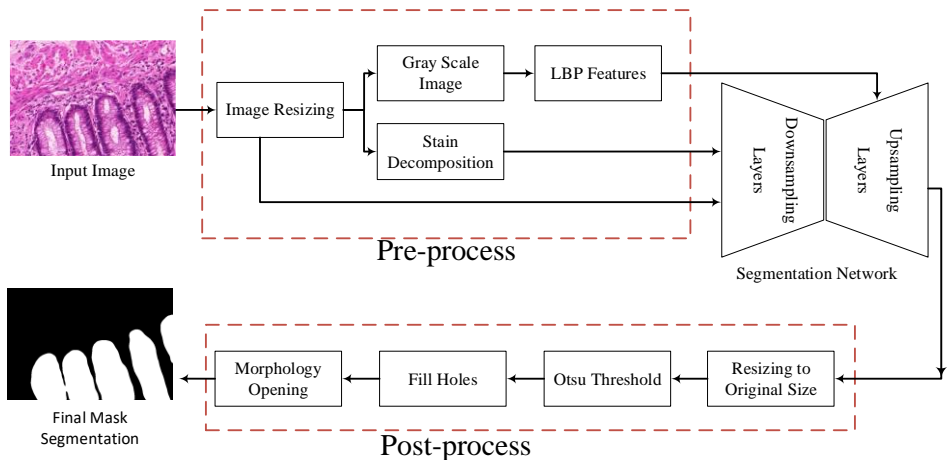

Figure 1. Block diagram of the gland segmentation system

stained images are decomposed to provide accurate and sharp color information for a comprehensive diagnosis. Wang *et al.* [8] use stain decomposition for classification of histopathological images where *H&E* components are picked as bilinear CNN inputs to enhance the visibility of spatial structures by color decomposition.

In this paper, we propose a deep network based on LinkNet structure for segmentation of glands. LinkNet [9] and U-Net [10] have been recently introduced for segmentation tasks. LinkNet is smaller with approximately one-third of the U-Net parameters and hence it is faster for training and test. Even though it is smaller, its performance is comparable to U-Net in terms of accuracy [9]. We show that the proposed structure can compete with the state-of-the-art methods in terms of Dice score, F1 score, and Hausdorff distance. The rest of this paper is structured as follows: technical details of the proposed system are described in section 2. Experimental results are discussed in section 3 and finally we conclude the paper and present possible future works' directions in section 4.

## II. PROPOSED METHOD

As shown in the block diagram of Figure 1, the proposed system is composed of two main modules for handling gland segmentation of histopathology images. A pre-processing module extracts appropriate input channels, namely the red channel and Hematoxylin component, as well as some handcrafted features. In some works RGB space is used for gland segmentation [3, 4, 6]. However, in this work we show that only red channel and the Hematoxylin component are more informative and lead to better segmentation. We modified LinkNet structure for incorporating specific features into the network structure. The proposed model is optimized through two loss functions with different entry points. In the last step, post-processing is applied to the probability map of the network output to improve the segmentation quality. We use the Otsu method [11] for estimating the best threshold and producing the segmentation binary mask from the probability map. Finally, morphological operations are applied in order to remove noise and small objects and fill the glands holes.

Fig.2 shows a sample image along with the binary mask and morphology improvements. Technical details of the other two system modules are described as follows.

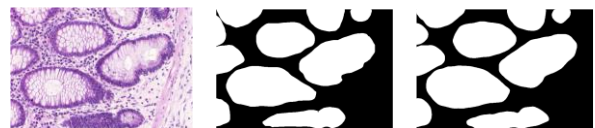

Figure 2. From left to right show the original image, ground truth and segmentation mask

### A. Preprocessing and Feature Selection

Since the size of images in our dataset is not uniform, we initially resize the input images to a fixed multiple of 64 pixels, both in width and height. Therefore, all images can be operated by the down-sampling layers of the segmentation network. In the second step, appropriate channels are extracted from the input image as well as handcrafted features to improve the network performance. Using RGB channels, for segmentation of the histopathology images, is a general approach in the research field. However, the boundary of glands in our dataset is more distinguishable in the red channel. In this work, we utilize two handcrafted features, namely invariant LBP features as well as *H&E* components, which are proved to be very informative for histopathology images [8]. For visual purposes, Figure 3 demonstrates a sample histopathology image along with the discussed channels/features. LBP features represent texture, and invariant LBP provides rotation invariant features for modeling the image texture [12].

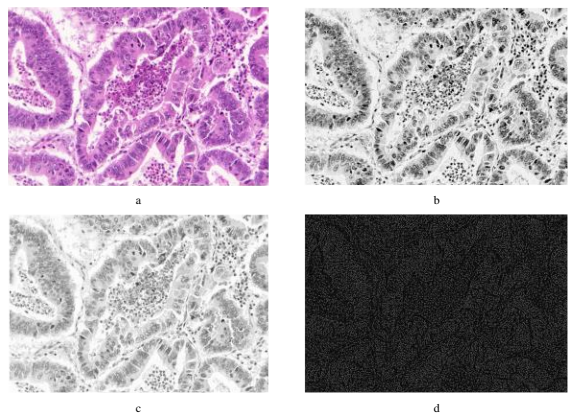

Figure 3. Sample preprocessing results from test B: a) original image, b) Hematoxylin component, c) Red channel and d) invariant LBP

Let's assume we have $N$ pixels in a neighborhood with radius $R$ and $LBP_{N,R}$ represents standard LBP features. Then the invariant LBP feature, as proposed in [11], is defined by the following equation:

$$LBP_{N,R}^{ir} = \min\{ROR(LBP_{N,R},i)|i = 0,1,\ldots,N-1\} \tag{1}$$

where $LBP_{N,R}^{ir}$ is the invariant LBP and $ROR$ refers to circular bitwise right shift. Fig.3-d demonstrates an example of invariant LBP features obtained from a gray-scale image.

Decomposition of $H\&E$ components is performed based on an orthonormal transformation of RGB space [12]. Each pure stain is characterized in various RGB channels with a different light absorption factor $c_i$. The ratio of incident and transmitted light, which is called optical density, is related to the amount of stain (A) multiplied by absorption factor $c_i$, as defined in the following equation:

$$I_i = I_{O,i} e^{-Ac_i} \tag{2}$$

where $I_{o,i}$ is the intensity of transmitted light after passing the specimen and $I_i$ is the intensity of the incident light in channel $i$. The optical density($OD$) is used to separate stain in each channel:

$$OD_i = -log_{10}\left(\frac{I_i}{I_{o,i}}\right) = kAc_i \; ; \; i \in \{R,G,B\} \tag{3}$$

For each channel, $OD$ is linearly related to the concentration of absorbing material ($c_i$), hence, it can be used for separation of the contribution of multiple stains in the specimen. Finally, $OD$ matrix of $M$ showing combination of three channels and three components Hematoxylin, Eosin and DAB is defined as:

$$M = \begin{bmatrix} 0.65 & 0.70 & 0.29 \\ 0.07 & 0.99 & 0.11 \\ 0.27 & 0.57 & 0.78 \end{bmatrix} \begin{matrix} \text{Hematoxylin} \\ \text{Eosin} \\ \text{DAB} \end{matrix}$$

If $C$ is the vector of absorption factors ($c_i$) for three stains at a particular pixel, then the vector of optical density levels detected at that pixel is $y = CM$ or equivalently $C = M^{-1}[y]$. We call $M^{-1}$ as the color deconvolution matrix and use it for calculation of absorption factors $C$.

We use this algorithm for separation of Hematoxylin, Eosin and DAB components.

### B. Segmentation Network

As discussed earlier, our proposed network is a modified version of LinkNet which is composed of encoder and decoder blocks. The shortcut connections between encoder and decoder layers of LinkNet structure help to maintain the resolution of spatial information and make it suitable for segmentation task [9].

A combination of Hematoxylin and red channel are fed into the network input layer as raw information. We also calculate the invariant LBP features from the grayscale image and inject them to the final layers of the network. To be more precise, LBP feature map is concatenated with the last deep features of the network prior to the segmentation layer.

We designate two output points for the network and define separate loss functions on them for training the network. The two outputs are supposed to estimate the segmentation probability map in different scales.

Hence, the middle layer tries to learn a coarse segmentation map and helps the last network layer to achieve higher performance. We use a weighted combination of two loss functions, with a higher emphasis on the coarse segmentation loss $((2 \times L_1) + L_2)$. Hence, the last layer is supposed to learn a fine-tuning over the coarse segmentation of earlier layers.

The loss function is defined by combining the binary cross entropy and Dice score. The cross-entropy is a pixel-based loss metric, while the Dice-score represents a holistic metric for training. Thus, the proposed combinatorial loss function is obtained from equations 4-6:

$$CE(G,I) = -(G \times log(I) + (1-G) \times log(1-I)) \tag{4}$$

$$D(G,I) = \frac{2 \times |G \cap I| + S}{|G| + |I| + S} \tag{5}$$

$$loss_{fun}(G,I) = CE(G,I) - e^{(1+D(G,I))} \tag{6}$$

Where G is the ground truth, I is the image and $S$ is a smoothing parameter ($S = 10^{-6}$).

### III. EXPERIMENTAL RESULT

The proposed gland segmentation system is implemented by python and tensor-flow. We use the public Warwick-QU dataset[1] [14] for training and evaluation. The dataset is composed of 85 color images for training and 80 test images in two separate sections (section A: 60 images and section B: 20 images). Since we do not use a pre-trained network, we have to augment our training dataset. The network training with 4420 augmented color images is conducted over 100 epochs. Training time is about 13 hours on NVIDIA GeForce GTX 1080 Ti. The proposed system is compared against two state-of-the-art researches [3], [7].

### A. Data Preparation

Since the 85 training images of Warwick-QU dataset have different sizes $\{(589 \times 453), (775 \times 522)\}$, we initially resize all of them to the closest multiple of 64 ($832 \times 576$) to be suitable for consecutive down-sampling layers of LinkNet.

For augmentation of training data, each image is rotated $180°$ and flipped in horizontal and vertical directions. Then, four overlapping large patches with ¾ of original size are cropped from the image corners and resized to $832 \times 576$. Finally, another round of $180°$ rotation, horizontal and vertical flipping is applied on all the images. Through this augmentation process, 4420 images are generated from the 85 original training images.

### B. Evaluation Metrics

Most researchers use the object-level Dice, F1 score and Hausdorff distance for evaluation of gland segmentation. Object level Dice is more suitable for individual objects than Dice metric [3]. In other words, when we have multiple glands in an image, object level Dice as defined below provides a more accurate evaluation on segmentation results:

$$D_{object}(G,I) = \frac{1}{2}\left(\sum_{i=1}^{n_I} w_i D(G_i,I_i) + \sum_{j=1}^{n_G} \widetilde{w_j} D(\widetilde{G_j},\widetilde{I_j})\right) \tag{7}$$

---

[1] http://www2.warwick.ac.uk/fac/sci/dcs/research/combi/research/bic/glascontest/

$I_i$ is the $i^{th}$ segmented gland and $G_i$ is a ground truth gland which has the maximum overlap with $I_i$. Similarly, $\widetilde{G}_j$ is the $j^{th}$ ground truth gland and $\widetilde{I}_j$ is the segmented gland with maximum overlap with $\widetilde{G}_j$. The weights $w_i$ and $\widetilde{w}_j$ are defined as $w_i = \frac{|I_i|}{\sum_{k=1}^{n_I}|I_k|}$ and $\widetilde{w}_j = \frac{|\widetilde{G}_j|}{\sum_{k=1}^{n_G}|\widetilde{G}_k|}$.

The object level Hausdorff distance represents the shape similarity, as defined below:

$$H(G,I) = max\left\{\sup_{x\in G}\inf_{y\in I}\|x-y\|, \sup_{y\in I}\inf_{x\in G}\|x-y\|\right\} \quad (8)$$

$$H_{object}(G,I) = \frac{1}{2}\left(\sum_{i=1}^{n_I} w_i H(G_i,I_i) + \sum_{j=1}^{n_G} \widetilde{w}_j H(\widetilde{G}_j,\widetilde{I}_j)\right) \quad (9)$$

F1 score is harmonic mean of recall and precision and is used for evaluation of gland detection:

$$F1 = \frac{2PR}{P+R}, P = \frac{TP}{TP+FP}, R = \frac{TP}{TP+FN} \quad (10)$$

where $TP$, $FP$, $FN$ stand for true positive, false positive and false negative respectively. For instance, each object detection is considered as true positive if the intersection between ground truth and the result is greater than 0.5. Otherwise, a false positive is reported. Similarly, FN is defined.

### C. Evaluation Results

We evaluate our system on two separate sections of Warwick-QU test set in gland segmentation challenge [14]. Quantitative results from sections A and B are presented in Table 1 and Table 2. The proposed system demonstrates competitive results in section B and comparable performance with the state-of-the-art in section A.

Table 1. Segmentation result on section B

| Method | Object Dice | F1 Score | Hausdorff |
| --- | --- | --- | --- |
| Xu et al. [7] | 0.815 | **0.771** | 129.930 |
| CUMedVision2 [3] | 0.781 | 0.716 | 160.347 |
| CUMedVision1 [3] | 0.800 | 0.769 | 153.646 |
| Our method | **0.822** | 0.750 | **108.208** |

The evaluation results of Table 1 demonstrate that our proposed system outperforms other methods on test B in terms of object Dice and Hausdorff distance. With regard to F1 score, our result is comparable with competitors.

Table 2. Segmentation result on section A

| Method | Object Dice | F1 Score | Hausdorff |
| --- | --- | --- | --- |
| Xu et al. [7] | 0.888 | 0.858 | 54.202 |
| CUMedVision2 [3] | **0.897** | **0.912** | **45.418** |
| CUMedVision1 [3] | 0.867 | 0.868 | 74.596 |
| Our method | 0.867 | 0.83 | 56.641 |

Table 2 demonstrates the evaluation results on test A. While our performance is not the best in this section, it is still comparable to competitors. As shown in the table, we have the same Object Dice with CUMedVision1 in Table 2, while we are better in Hausdorff. CUMedVision2 utilize an additional contour based tool which helps them discriminate between the very close gland objects of section A by avoiding to connect the gland borders. This additional tool helps them to outperform others in test A, while all the results are still competing very closely. Even though we are not using any additional tool to help our segmentation network, experimental results of Table 1 show that we outperform CUMedVision1 and CUMedVision2 in terms of Object Dice and Hausdorff distance.

### IV. CONCLUSION

In this paper, we presented a modified version of LinkNet for gland segmentation in histopathology images. We used the Red channel and Hematoxylin component as network inputs. Utilizing Invariant LBP features for modeling image texture and feeding them into the final network layers have increased the system performance. Furthermore, the employment of the double loss functions in different scales has improved the performance of the gland segmentation. This mechanism lets the network fine-tune a coarse segmentation map along the final network layers.